\documentclass{PoS}

\title{Glueballs in charmonia radiative decays}

\ShortTitle{Glueballs in charmonia radiative decays}

\author{Ying Chen\\
        Institute of High Energy Physics, Chinese Academy of Sciences, Beijing 100049 China\\
        E-mail: \email{cheny@ihep.ac.cn}}
\author{Long-Cheng Gui\\
        Institute of High Energy Physics, Chinese Academy of Sciences, Beijing 100049 China\\
        E-mail: \email{guilongcheng@ihep.ac.cn}}
\author{Gang Li\\
        Department of Physics, Qufu Normal University, Qufu 273165 China\\
        E-mail: \email{gli@mail.qfnu.edu.cn}}
\author{\speaker{Chuan Liu}
     \\
        School of Physics and Center for High Energy Physics, Peking University, Beijing 100871, China\\
        E-mail: \email{liuchuan@pku.edu.cn}}
\author{Yu-Bin Liu\\
        School of Physics, Nankai University, Tianjin 300071 China\\
        E-mail: \email{liuyb@nankai.edu.cn}}
\author{Jian-Ping Ma\\
        Institute of Theoretical Physics, Chinese Academy of Sciences, Beijing 100190 China\\
        E-mail: \email{majp@itp.ac.cn}}
\author{Yi-Bo Yang\\
        Institute of High Energy Physics, Chinese Academy of Sciences, Beijing 100049 China\\
        E-mail: \email{yangyb@ihep.ac.cn}}

\author{Jian-Bo Zhang\\
        Department of Physics, Zhejiang University, Hangzhou 311027 China\\
        E-mail: \email{jbzhang08@zju.edu.cn}}

\abstract{ Scalar~\cite{scalar_paper} and tensor~\cite{tensor_paper} glueballs
 created in $J/\psi$ radiative decays are studied in quenched lattice QCD.
 Using two anisotropic lattices to approach the continuum limit,
 we compute the relevant form factors responsible for
 the decay rates for $J/\psi\rightarrow\gamma G_{0^{++}}$ and $J/\psi\rightarrow\gamma G_{2^{++}}$.
 Comparing with the existing experimental data, it is argued that $f_0(1710)$ is a favorable
 candidate for scalar glueball. The decay rate for $J/\psi\rightarrow\gamma G_{2^{++}}$ is found to
 be quite substantial. A comprehensive search in the tensor channel on BESIII is therefore suggested.}

\FullConference{31st International Symposium on Lattice Field Theory - LATTICE 2013\\
		July 29 - August 3, 2013\\
		Mainz, Germany}

\def \be {\begin{equation}}
\def \ee {\end{equation}}
\def \ba {\begin{eqnarray}}
\def \ea {\end{eqnarray}}
\begin{document}

\section{Introduction}

 Glueballs are exotic hadronic states made up of gluons. Their existence is permitted by QCD but remains
 to be confirmed by experiments. Quenched lattice QCD studies~\cite{prd56,prd60,prd73} have been performed and the mass value for the scalar and tensor glueball turns out to be around 1.7GeV and 2.4GeV respectively.
 Recent exploratory study suggests that the situation might be similar for dynamical fermions~\cite{Gregory:2012},
 as far as the mass values are concerned.

 It is well-known that gluons can be copiously produced in $J/\psi$ radiative decays.
 It is expected that the gluons produced in $J/\psi$ radiative decays dominantly form a glueball.
 If the production rate of the glueball in the radiative
 decay can be obtained from theoretical studies, it will provide important information for identifying the
 possible candidate for the glueballs.
 Due to its obvious non-perturbative nature, lattice QCD is the choice for this study
 from first principles. In this paper, we investigate the radiative
 decay of $J/\psi$ into a scalar or a tensor glueball in quenched lattice QCD~\cite{scalar_paper,tensor_paper}.
 Our results will shed some light on various issues concerning the glueball candidates that
 have been searched for at BEPCII with by far the largest $J/\Psi$ sample in the world.

\section{Lattice setup}

 To lowest order in QED, the amplitude for decay $J/\psi\rightarrow \gamma G$ is given by
 \begin{equation}
 M_{r,r_\gamma,r_G}=\epsilon_{\mu}^*(\vec{q},r_\gamma)\langle
 G(\vec{p}_f,r_G)|j^{\mu}(0)|J/\psi(\vec{p}_i,r)\rangle,
 \end{equation}
 where $\vec{p}_i$ is the initial three-momentum of $J/\Psi$
 while $\vec{p}_f$ is the final momentum of glueball $G$;
 $r$, $r_\gamma$ and $r_G$ corresponds to the helicity index
 for the $J/\Psi$, photon and the glueball, respectively.
 We use $\vec{q}=\vec{p}_i-\vec{p}_f$  to designates the
 three-momentum of the real photon with $\epsilon(\vec{q},r_\gamma)$ being its polarization vector.
 The electromagnetic current operator is given by: $j^\mu=\sum_f Q_f\bar{q}_f\gamma^\mu q$ with $Q_f$ being
 the electric charge for flavor $f$.

 It turns out that matrix element $\langle G(\vec{p}_f,r_G)|j^{\mu}(0)|J/\psi(\vec{p}_i,r)\rangle$,
 which is non-perturbative in nature, can be related to the following three-point functions, see
 e.g. Ref.~\cite{dudek06,Niu:2011}, that are computable in lattice QCD:
 \begin{eqnarray}
 \label{eq:three-point-function-def}
&&\!\!\!\!\!\!\!\!\!\!\!\!\!\!\!\!\!\!\!\!\!\!\!\!\!\!\!\!\Gamma_{i,\mu,j}^{(3)}(\vec{p}_f,\vec{q};t_f,t) = \frac{1}{T}\sum\limits_{\tau=0}^{T-1}\sum\limits_{\vec{y}}
e^{-i\vec{q}\cdot \vec{y}} \langle
\Phi^{(i)}(\vec{p}_f,t_f+\tau)J_\mu (\vec{y},t+\tau)O_{V,j}(\vec{0},\tau)\rangle\;, \\
\label{eq:three-point-function-propagators}
&& \!\!\!\!\!\!\!\!\!\!\!\!\!\!=\frac{1}{T}\sum\limits_{\vec{y},\tau=0}^{T-1} e^{-i\vec{q}\cdot \vec{y}} \left\langle
\Phi^{(i)}(\vec{p}_f,t_f+\tau){\rm Tr}\left[\gamma_\mu S_F(\vec{y},t+\tau;\vec{0},\tau)\gamma_j\gamma_5
S_F^\dagger(\vec{y},t+\tau;\vec{0},\tau)\gamma_5\right]\right\rangle\;,\\
 \label{eq:three-point-function-completeset}
&&\!\!\!\!\!\!\!\!\!\!\!\!\!\!=\!\!\!\sum\limits_{G,V} \frac{e^{-E_G (t_f-t)}e^{-E_V
t}}{2E_G(\vec{p}_f)V_3 2E_V(\vec{p}_i)} \langle
0|\Phi^{(i)}(0)|G(\vec{p}_f)\rangle \langle
G(\vec{p}_f)|J_\mu(0)|V(\vec{p}_i)\rangle\langle
V(\vec{p}_i)|O_{V,j}^{\dagger}(0)|0\rangle\;.
\end{eqnarray}
 In the first of these expressions, $O_{V,j}$ is the operator which creates a vector charmonium from the QCD
 vacuum while $\Phi^{(i)}$ is the optimized pure gauge glueball operator that is
 obtained from a variational computation in the pure gauge sector~\cite{prd56,prd60,prd73}.
 The operator $J_\mu(x)=\bar{c}\gamma_\mu c(x)$ is the vector current of the charm quark.
 Since disconnected and OZI-suppressed contributions are neglected in this computation,
 contribution from other quark flavors are dropped out. Note, however, this type of vector current
 is not conserved on the lattice and requires an extra multiplicative renormalization factor $ Z^{(s)}_V(a_s)$ which is
 computed non-perturbatively using the spatial components of the current in our study~\cite{scalar_paper}. Furthermore, as
 Eq.~(\ref{eq:three-point-function-propagators}) indicates, the connected contributions from
 the charm quark can further be expressed in terms of charm quark propagators.
 When inserting a complete set of states in between the above mentioned operators,
 it is realized that the three-point function in Eq.~(\ref{eq:three-point-function-def})
 becomes a sum over all possible contributions from intermediate
 states, i.e. the sum over $G$ and $V$ in Eq.~(\ref{eq:three-point-function-completeset}).
 The energies $E_G(\vec{p}_f)$ and $E_V(\vec{p}_i)$ and the overlap matrix elements
  $\langle 0|\Phi^{(i)}(0)|G(\vec{p}_f)\rangle$ and $\langle V(\vec{p}_i)|O_{V,j}^{\dagger}(0)|0\rangle$
  can be obtained from the corresponding two-point functions for the
  operator $\Phi^{(i)}(0)$ and $O_{V,j}(0)$, respectively.
 In a scenario $t_f\gg t\gg 1$, the three-point function is dominated by the ground state contribution
 which contains the desired matrix element $\langle G(\vec{p}_f)|J_{\mu}(0)|J/\psi(\vec{p}_i)\rangle$
 that we are after. In real simulations, one could design appropriate ratios of
 three-point functions and two-point functions such that a plateau behavior in $t$
 yields the corresponding matrix element $\langle G(\vec{p}_f)|J_{\mu}(0)|J/\psi(\vec{p}_i)\rangle$.
 For example, for the case of the tensor glueball, one forms the following ratio,
 \footnote{In this calculation, we take the reference frame such that the tensor glueball is at rest.}
 \be
    R_{i,\mu,j}(\vec{q},t)=\Gamma^{(3)}_{i,\mu,j}(\vec{q},t_f,t)\frac{\sqrt{4V_3 M_TE_V(\vec{q})}}{C^{i}(t_f-t)}\sqrt{\frac{\Gamma^{(2)}_j(\vec{q},t_f-t)}
    {\Gamma^{(2)}_j(\vec{q},t)\Gamma^{(2)}_j(\vec{q},t_f)}}.
 \ee
 Here $\Gamma^{(2)}_j(\vec{q},t)$ is the two-point correlation function
 for the $J/\Psi$ operator $O_{V,j}$ while $C^{i}(t)$ is the glueball
 two-point function for the optimized glueball operator $\Phi^{(i)}$.
 With the relevant factors obtained from corresponding two-point functions and
 by searching for plateau behavior in $t$ for various values of $Q^2$, this
 ratio $R_{i,\mu,j}(\vec{q},t)$ gives us the desired hadronic matrix element,
 \begin{equation}
R_{i,\mu,j}(\vec{q},t)=\sum_{r}\langle
T_i|J_{\mu}(0)|V(\vec{q},r)\rangle\epsilon_j(\vec{q},r)+\delta f(t),
\end{equation}
 where $\epsilon_j(\vec{q},r)$ is the polarization vector for $J/\Psi$ and
 $\delta f(t)$ accounts for the contaminations from excited states.

 In the continuum limit, the matrix element that we obtain from the lattice can be decomposed into
 appropriate form factors. For example, for the case of the scalar glueball, we have
   \begin{equation}
\sum_r \langle S(\vec{p}_f)|J_\mu(0)|V(\vec{p}_i,r)\rangle
\epsilon_j(\vec{p}_i,r)=\alpha_{\mu j}E_1(Q^2)+\beta_{\mu j}C_1(Q^2),
 \end{equation}
 where $E_1(Q^2)$ and $C_1(Q^2)$ are the corresponding form factors which are functions
 of the photon four-momentum squared $Q^2=-(p_f-p_i)^2$. Factors $\alpha_{\mu j}$
 and $\beta_{\mu j}$ are known kinematic functions of initial and final momenta.
 Similarly for the case of tensor glueball, we have
 \begin{equation}
\!\!\!\!\!\langle G(\vec{p}_f, r_G) | J_{\mu}(0) |V(\vec{p}_i,r)\rangle = \alpha_1^\mu
E_1(Q^2) + \alpha_2^{\mu}M_2(Q^2) + \alpha_3^\mu E_3(Q^2) + \alpha_4^\mu C_1(Q^2)+ \alpha_5^\mu C_2(Q^2).
\end{equation}
 Again, $E_1(Q^2)$, $M_2(Q^2)$, $E_3(Q^2)$, $C_1(Q^2)$ and $C_2(Q^2)$ are
 the corresponding form factors while $\alpha^\mu_i$'s are known kinematic functions,
 see e.g. Ref.~\cite{Yang:2012}.

 For the physical decay width, one has to take the
 form factors evaluated at the physical photon point $Q^2=0$.
 Thus, for the case of scalar and tensor glueballs, we have
  \begin{eqnarray}
  \label{eq:scalar_width}
\Gamma(J/\psi\rightarrow \gamma
G_{0^{++}})&=&\frac{4\alpha|\vec{p}_\gamma|}{27M_{J/\psi}^2}|E_1(0)|^2,
 \\
 \label{eq:tensor_width}
 \Gamma(J/\psi\rightarrow \gamma
G_{2^{++}})&=&\frac{4\alpha|\vec{p}_\gamma|}{27M_{J/\psi}^2}\left(|E_1(0)|^2+|M_2(0)|^2+|E_3(0)|^3\right),
 \end{eqnarray}

\section{Numerical results}

 This calculation was performed on anisotropic lattices~\cite{prd56} using tadpole improved Wilson
 fermions~\cite{chuan1}. The bare anisotropy parameter is set to $\xi=a_s/a_t=5$ which greatly enhanced
 the resolution in the temporal direction. Two different spatial lattice spacings
 have been used, the coarse lattice with $a_s=0.222$fm ($\beta=2.4$) and
 the fine lattice with $a_s=0.138$fm ($\beta=2.8$)
 to inspect possible lattice spacing errors, where $a_s$ values are determined from $r_0^{-1}=410(20)$ MeV.
 The parameters in the action are tuned carefully by requiring that the physical dispersion
 relations of vector and pseudoscalar mesons are correctly reproduced at each bare quark mass~\cite{chuan2}. The bare
 charm quark masses at different $\beta$ are determined by the physical mass of $J/\psi$, $m_{J/\psi}=3.097$ GeV.
 Relevant input parameters are summarized in Table~\ref{tab:lattice}.
\begin{table}[htb]
\centering \caption{\label{tab:lattice} The input parameters for the calculation. Values for the
coupling $\beta$, anisotropy $\xi$, the lattice spacing $a_s$, lattice size, and the number of
measurements are listed.}
\begin{tabular}{cccccc}
 \hline
 $\beta$ &  $\xi$  & $a_s$(fm) & $La_s$(fm)&
 $L^3\times T$ & $N_{\rm conf}$ \\\hline
   2.4  & 5 & 0.222(2) & 1.78 &$8^3\times 96$ & 5000 \\
   2.8  & 5 & 0.138(1) & 1.66 &$12^3\times 144$ & 5000  \\
   \hline
\end{tabular}
\end{table}
 Another trick that have been utilized is the average over temporal time-slices
 which effectively increased our statistics, see e.g. Eq.~(\ref{eq:three-point-function-def}).

 In the data analysis, the 5000 configurations are divided into 100 bins and the average of 50
 measurements in each bin is taken as an independent measurement. For the resultant 100
 measurements, the one-eliminating jackknife method is used to perform the fit for the matrix
 elements. Since the matrix elements are measured from the same
 configuration ensemble, we carry out a correlated data fitting to get the form factors
 simultaneously with covariance matrix constructed from the jackknife method.

 \begin{figure}[htb]
\includegraphics[height=5cm]{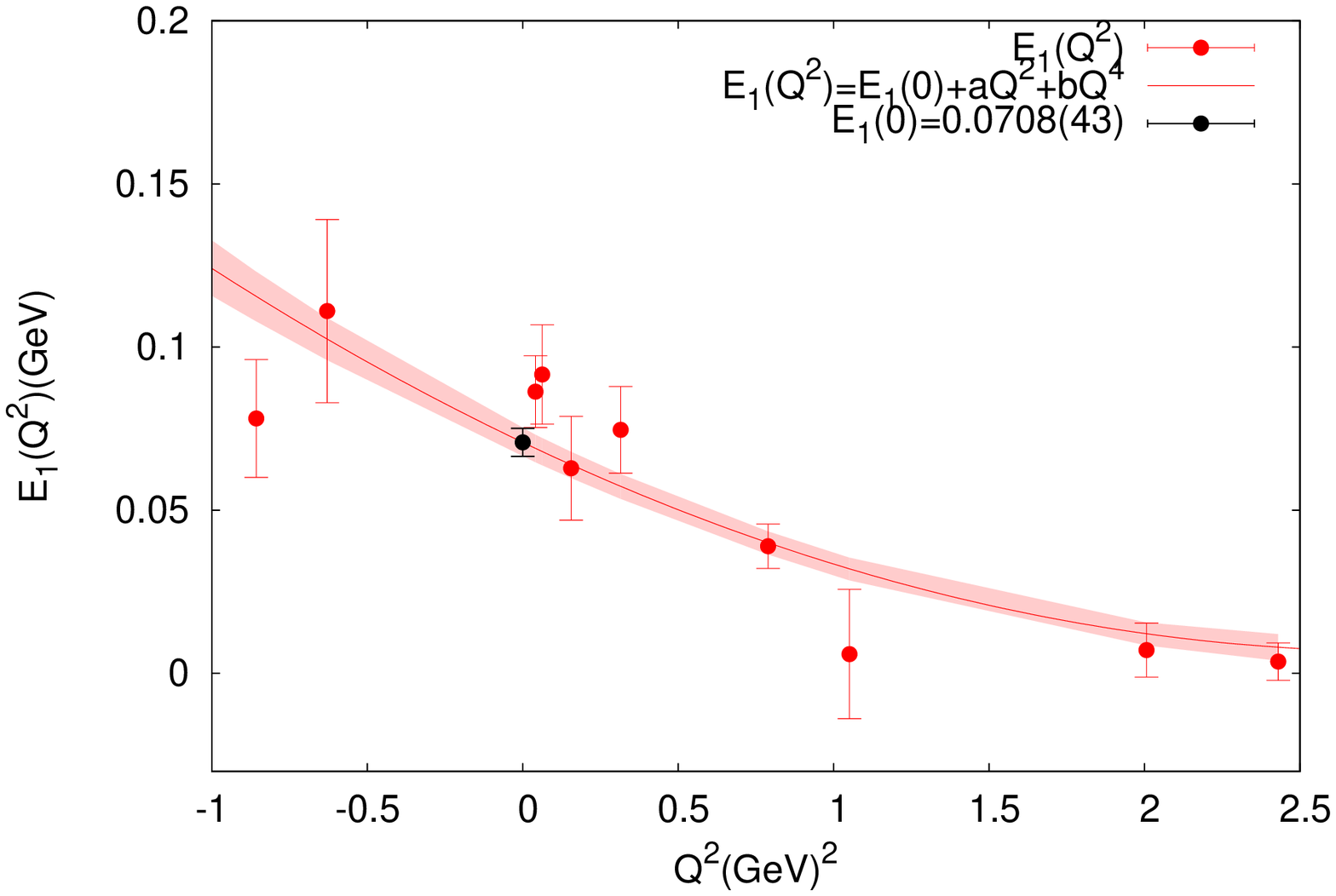}
\includegraphics[height=5cm]{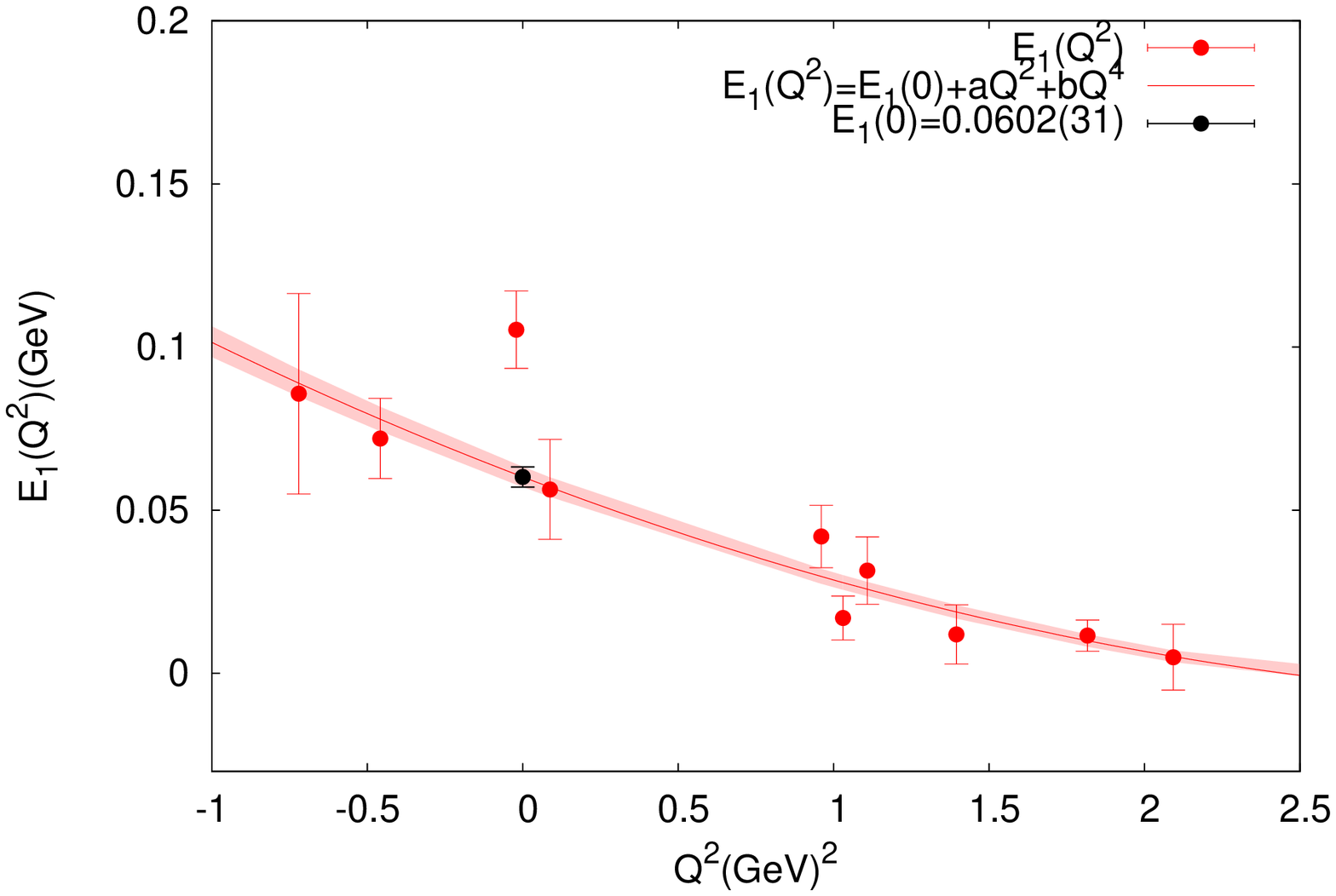}
\caption{\label{fig:form_scalar}
The extracted form factors $E_1(Q^2)$ in physical units. The left panel is for $\beta=2.4$
and the right one for $\beta=2.8$. The curves with error bands indicate the polynomial fit with
$E_1(Q^2)=E_1(0)+aQ^2+bQ^4$ while the black dot being the interpolated value $E_1(0)$ at $Q^2=0$.}
\end{figure}
 For the scalar glueball, only one form factor, namely $E_1(Q^2=0)$, is relevant for the decay.
 After obtaining the form factor $E_1(Q^2)$ at various values of $Q^2$, we fit the
  form factor from $Q^2=-1.0 \,{\rm GeV}^2$ to $2.5 \,{\rm GeV}^2$ using a polynomial form:
  \be
  \label{eq:fit_scalar}
  E_1(Q^2)=E_1(0)+aQ^2+bQ^4\;.
  \ee
  This is done for both the coarse ($\beta=2.4$) and the fine ($\beta=2.8$) lattice.
  In Fig.~\ref{fig:form_scalar}, we show the form factor $E_1(Q^2)$ obtained from our lattice calculations at
 the two lattice spacings. The left/right panel corresponds to the coarse/fine lattice, respectively.
 The data points are indicated by the red solid points while the shaded bands designate the
 polynomial fit~(\ref{eq:fit_scalar}). The fitted values for $E_1(0)$ are shown by the black solid
 points at $Q^2=0$ in each panel. This particular value is to be substituted into
 Eq.~(\ref{eq:scalar_width}) for the decay width of $J/\psi\rightarrow \gamma G_{0^{++}}$.

\begin{figure}[htb]
\includegraphics[height=5cm]{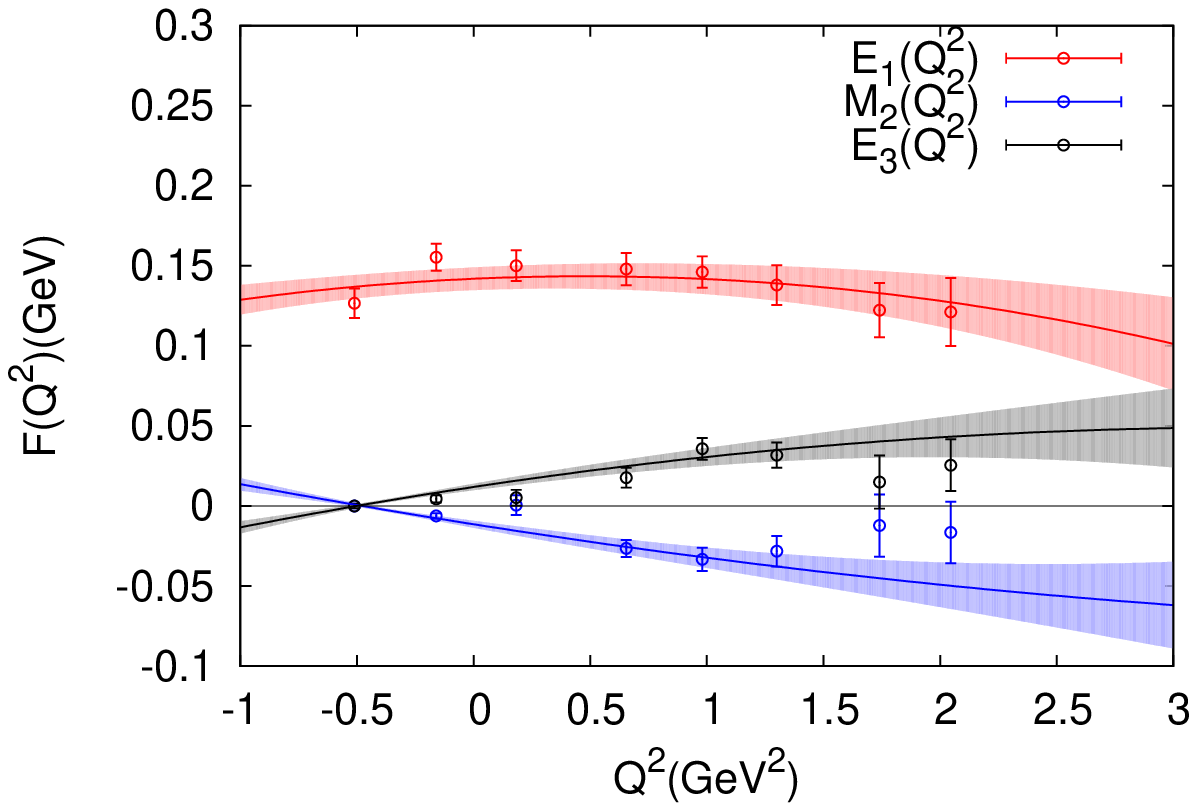}
\includegraphics[height=5cm]{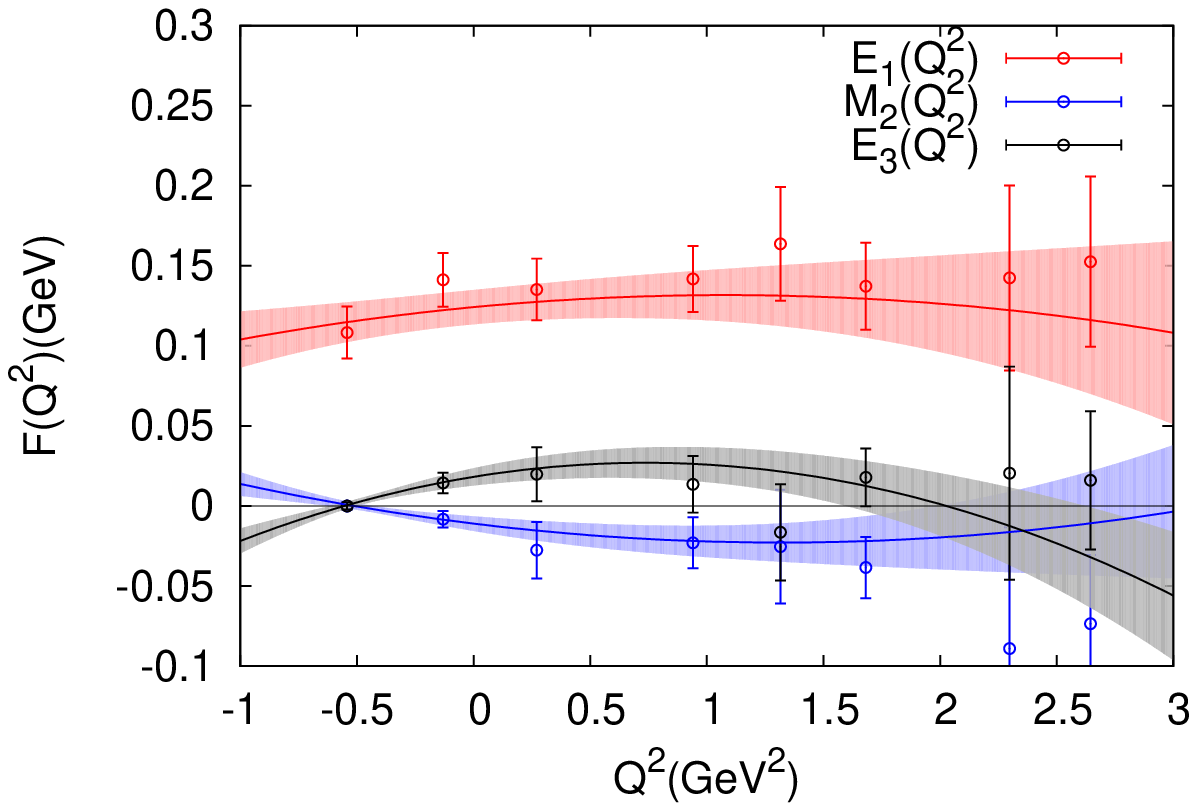}
\caption{\label{fig:form_tensor}The extracted form factors $E_1(Q^2)$ $M_2(Q^2)$ and $E_3(Q^2)$ in the
physical units. The left panel is for $\beta=2.4$ and the right one for $\beta=2.8$. The curves
with error bands show the polynomial fit with $F_i(Q^2)=F_i(0)+a_iQ^2+b_iQ^4$.}
\end{figure}
 For the tensor glueball, the analysis is similar except that we have
 three form factors: $E_1$, $M_2$ and $E_3$. The situation is shown
 in Fig.~\ref{fig:form_tensor} with the left/right panel corresponds
 to the coarse/fine lattice. Again, these form factors are
 fitted from $Q^2=-0.5 \,{\rm GeV}^2$ to $2.7\,{\rm GeV}^2$ using polynomials,
   \be
  \label{eq:fit_tensor}
  F_i(Q^2)=F_i(0)+a_iQ^2+b_iQ^4\;,
  \ee
  with $i=1,2,3$ corresponds to $E_1$, $M_1$ and $E_3$, respectively.
 More sophisticated fitting forms and different fitting ranges have also been attempted,
 however, statistical compatible results was obtained.

 Results for the form factors
 obtained thus far, together with the corresponding glueball mass values for the scalar and tensor,
 are summarized in Table~\ref{tab:summary}. Also listed in the Table are the renormalization
 factor $Z_V^{(s)}$ for the two lattices and the decay width computed using
 Eq.~(\ref{eq:tensor_width}). With these values at finite lattice spacing,
 one could perform a linear extrapolation in $a^2_s$ to estimate the finite
 lattice spacing errors. These extrapolated values are also listed where applicable.
\begin{table}[htb]
\centering \caption{\label{tab:summary} Results for scalar and tensor glueballs.}
\begin{tabular}{ccccc}
 \hline
 $\beta$  &  $M_{0^{++}}$(GeV) & $Z_V^{(s)}(a_s)$  &  $E_1(0,a_s)$ (GeV) & $\Gamma$(keV)  \\
 \hline
   2.4   &  1.360(9)   & 1.39(2)   &    0.0708(43)     &    \ldots    \\
   2.8   &  1.537(7)   & 1.11(1)   &    0.0602(31)     &    \ldots    \\
   $\infty$ & 1.710(90)& \ldots       &    0.0536(57)     &    0.35(8) \\
\end{tabular}
\begin{tabular}{cccccc}
 \hline
 $\beta$  &  $M_{2^{++}}$(GeV)  &  $E_1$ (GeV) & $M_2$ (GeV) &  $E_3$ (GeV) & $\Gamma$(keV)  \\
 \hline
   2.4   &  2.360(20)     &    0.142(07)       &  -0.012(2)  & 0.012(2) & 1.46(18)    \\
   2.8   &  2.367(25)     &    0.125(10)       &  -0.011(4)  & 0.019(6) & 1.17(20)    \\
   $\infty$ & 2.39(12)    & 0.114(12)          &  -0.011(5)  & 0.023(8) & 0.99(22)    \\
 \hline
\end{tabular}
\end{table}

 We now turn to phenomenological implications of our results. First the scalar glueball
 case. As is known, there are three major candidates in the scalar channel: $f_0(1370)$,
 $f_0(1500)$ and the $f_0(1710)$. Our lattice result shows that the branching ratio
 \begin{equation}
 \label{eq:br_scalar}
 \Gamma(J/\psi\rightarrow \gamma G_{0^{++}})/\Gamma_{\rm tot}=3.8(9)\times 10^{-3}.
 \end{equation}
 Although the final states measured in experiments are not pure gauge glueballs,
 this branching ratio can give us useful information about which of the
 three candidates, $f_0(1710)$, $f_0(1500)$ and $f_0(1370)$,
 has a larger pure gauge glueball component. By comparing with
 the existing experimental data, we concluded that~\cite{scalar_paper} only $f_0(1710)$ is
 compatible with the branching ratio~(\ref{eq:br_scalar}), making it
 the dominant candidate for the scalar glueball. At least, we could say that
 $f_0(1710)$ contains a much more substantial glueball component than the
 other two candidates.

 For the case of tensor glueball, our lattice result indicates a large
 branching ratio,
 \begin{equation}
\Gamma(J/\psi\rightarrow \gamma G_{2^{++}})/\Gamma_{\rm tot}=1.1(2)\times 10^{-2}.
\end{equation}
 With such a large branching ratio, tensor glueballs should have been created
 abundantly  in $J/\Psi$ radiative decays. However, there is no obvious candidates
 experimentally observed so far. The narrow state $f_J(2220)$ observed by Mark III and BES in the $J/\psi$ decay was
 once interpreted as a candidate for the tensor glueball. Nevertheless,
 BESII with substantially more statistics does not find the evidence of a
 narrow structure around $2.2\,{\rm GeV}$ of $\pi\pi$ invariant mass spectrum in the processes
 $J/\psi\rightarrow \gamma \pi\pi$~\cite{Ablikim06}. Recently, based on 225 million $J/\psi$ events,
 the BESIII Collaboration performs a partial wave analysis of $J/\psi\rightarrow \gamma\eta\eta$ and
 also finds no evident narrow peak for $f_J(2220)$ in the $\eta\eta$ mass
 spectrum~\cite{Ablikim:2012}. So the existence of $f_J(2220)$ is still very weak.
 It is possible that tensor glueball in this mass
 range mix with the other hadronic final states strongly such that no single channel
 is dominant. Our result thus  motivates a serious joint analysis of the radiative $J/\psi$ decay into tensor
 objects in $VV$, $PP$, $p\bar{p}$ and $4\pi$ final states (where $V$ and $P$ stand for vector and
 pseudoscalar mesons, respectively), among which $VV$ channels may be of special importance since
 they are kinematically favored in the decay of a tensor meson.

\section{Conclusions}

 Glueballs are supposed to be produced copiously in charmonia radiative decays.
 BESIII, with the largest charmonia sample in the world, seems to be the best hunting ground for glueballs.
 In this exploratory quenched lattice study, we computed the radiative transition rate
 of $J/\Psi$ to scalar and tensor glueballs. Our calculation suggests that
 $f_0(1710)$ contains more scalar glueball components than
  other candidates like $f_0(1500)$ and $f_0(1370)$.
 Our results also indicate that the radiative decay rate for tensor glueball is quite large.
 A comprehensive search is suggested in the tensor channels at BESIII to
 gain further information about tensor glueballs.
 Finally, unquenched lattice study is very much welcome which will clarify a lot of
 remaining puzzles.

\section*{Acknowledgments}

 This work is supported in part by the
 National Science Foundation of China (NSFC) under the project
 No. 11335001, No. 11075167, No.11021092, No. 11275169 and No. 10975076.
 It is also supported in part by the DFG and the NSFC (No.11261130311) through funds
 provided to the Sino-Germen CRC 110 ``Symmetries and the Emergence of Structure in QCD''.


\begin{thebibliography}{99}
 \bibitem{scalar_paper} Long-Cheng~Gui {\it et al.}, (CLQCD Collaboration), Phys.Rev.Lett. 110, 021601 (2013)
 \bibitem{tensor_paper} Yi-Bo~Yang {\it et al.}, (CLQCD Collaboration), Phys. Rev. Lett. 111, 091601 (2013) 
 \bibitem{prd56} C.J. Morningstar and M. Peardon, Phys. Rev. D {\bf 56}, 4043 (1997).
 \bibitem{prd60} C.J. Morningstar and M. Peardon, Phys. Rev. D {\bf 60}, 034509 (1999).
 \bibitem{prd73} Y.~Chen {\it et al.}, Phys. Rev. D {\bf 73}, 014516 (2006).
  \bibitem{Gregory:2012} E.~Gregory et al, JHEP 10, 170 (2012)
\bibitem{dudek06} J.J. Dudek, R.G. Edwards, and D.G. Richards, Phys. Rev. D {\bf 73}, 074507
(2006).
\bibitem{Niu:2011} Y.~Chen, {\it et al.}, (CLQCD Collaboration), Phys. Rev. D {\bf 84}, 034503 (2011).
\bibitem{Yang:2012} Y.-B. Yang, Y. Chen, L.-C. Gui, C. Liu, Y.-B. Liu, Z. Liu, J.-P. Ma, and J.-B. Zhang (CLQCD
Collaboration), Phys. Rev. D {\bf 87}, 014501 (2013).
 \bibitem{chuan1}C. Liu, J. Zhang, Y. Chen, J.P. Ma,
Nucl. Phys. B {\bf 624}, 360 (2002).
\bibitem{chuan2} S. Su, L. Liu, X. Li, and C. Liu, Int. J. Mod.
Phys. A {\bf 21}, 1015 (2006), Chin. Phys. Lett. {\bf 22}, 2198 (2005).
 \bibitem{Ablikim06} M. Ablikim {\it et al.}(BES Collaboration), Phys. Lett. B {\bf 642}, 441
(2006).
\bibitem{Ablikim:2012} M. Ablikim {\it et al.}(BES Collaboration), arXiv:1301.0053 (hep-ex).

\end{thebibliography}
\end{document}